\documentclass[12pt]{article}

\input epsf
\def\ba{\begin{eqnarray}}
\def\ea{\end{eqnarray}}
\def\beq{\begin{eqnarray}}
\def\eeq{\end{eqnarray}}
\def\mpl{M_{\rm Pl}}
\def\t g{\tilde g}
\def\({\left(}
\def\){\right)}

\def\lsim{\mathrel{\rlap{\lower3pt\hbox{\hskip0pt$\sim$}}
     \raise1pt\hbox{$<$}}}         
\def\gsim{\mathrel{\rlap{\lower4pt\hbox{\hskip1pt$\sim$}}
     \raise1pt\hbox{$>$}}}         

\usepackage{amsmath}
\usepackage{amsfonts}

\begin{document}

\begin{titlepage}

\begin{flushright}
{NYU-TH-06/12/10}\\
\end{flushright}
\vskip 0.9cm

\centerline{\Large \bf Selftuned  Massive Spin-2}
\vskip 0.7cm
\centerline{\large Claudia de Rham\,$^1$ and Gregory Gabadadze\,$^2$}
\vskip 0.3cm

\centerline{\em $^1$\,D\'epartment de Physique  Th\'eorique, Universit\'e
de  Gen\`eve,}
\centerline{\em 24 Quai E. Ansermet, CH-1211,  Gen\`eve, Switzerland}

\centerline{\em $^2$\,Center for Cosmology and Particle Physics,
Department of Physics, }
\centerline{\em New York University, New York,
NY, 10003, USA}

\vskip 1.9cm

\begin{abstract}

We calculate the cubic order terms in a covariant theory that
gives a nonlinear completion of the Fierz-Pauli massive spin-2 action.
The resulting terms have  specially tuned  coefficients
guarantying the absence of a ghost at this order in the decoupling limit.
We show in this limit that:
(1) The quadratic theory propagates helicity-2, 1, and helicity-0
states of massive spin-2.
(2) The  cubic terms with six derivatives
-- which would give ghosts on local backgrounds --
cancel out automatically.
(3) There is a four-derivative cubic term for the helicity-0 field,
that has been known to be ghost-free on any local  background.
(4) There are  four-derivative cubic terms  that mix two helicity-0 fields
with one helicity-2, or two helicity-1 fields  with one helicity-0;
none of them give ghosts on local backgrounds.
(5) In the absence of external  sources,   all the  cubic mixing terms
can be removed by nonlinear redefinitions of  the helicity-2 and
helicity-1 fields.  Notably, the helicity-2 redefinition
generates the  quartic Galileon term. These findings hint
to an underlying nonlinearly realized symmetry,
that should be responsible for what appears as
the accidental cancellation of the ghost.

\end{abstract}

\vspace{3cm}

\end{titlepage}

\newpage

\subsection*{1. Introduction and summary}

An effective  field theory for  massive spin-2 is motivated by
spin-2 QCD resonances (glueballs, quark-antiquark  mesons,  or their
mixtures) which become long-lived in the  limit of a large 
number of colors.  It is also motivated by massive gravity, and
a possibility of having dark energy made of ``condensate'' of
massive gravitons.  In what follows we will be discussing a classical
theory of massive gravity on asymptotically flat space-time.

The Fierz-Pauli (FP) Lagrangian \cite{FP} is a linear
theory describing a massive spin-2  state on flat space,
without any ghosts or tachyons  \cite{Nieu}.
Recently, a nonlinear completion to the
FP theory  to all orders was proposed in Refs. \cite{GG,CdR}.
The gravitational  field in this approach
is described by an extended  metric tensor
${\t g}_{\mu\nu}(x,u)$, with $\mu,\nu=0,1,2,3,$
which is labeled by a continuous dimensionless parameter $u$.
The matter fields  do not depend on $u$,
but couple to the metric tensor $g_{\mu\nu}(x)\equiv
{\t g}_{\mu\nu}(x,u=0).$ The Lagrangian density for the
gravitational field alone reads as follows:
\beq
{\cal L}= {\mpl^2} \sqrt{g} R -
{\mpl^2 m^2\over 2}\int_{-1}^{+1}du \sqrt{ \t g}
\left (k_{\mu\nu}^2 - k^2 \) \,,
\label{egr}
\eeq
where $R$ is the Ricci scalar of the metric $g_{\mu\nu}(x)$, while
$k_{\mu\nu}\equiv{1\over 2}\partial_u \t g_{\mu\nu}$,
$k \equiv \t g^{\mu\nu} k_{\mu\nu} $;  all
indices in the Einstein-Hilbert term in (\ref {egr}) are raised
by $g^{\mu\nu} $,  while those in the second term
by $\t g^{\mu\nu}$. The ${\bf Z_2}$  symmetry is imposed  on the fields,
$ {\t g}_{\mu\nu}(x,u) = \t g_{\mu\nu}(x,- u)$.
The ``$u$-dimension'' is not dynamical since
fields have no ordinary derivative terms for $u>0$,
and there are no $g_{uu}$ or $g_{u\mu}$ components of the metric
to vary.

\vspace{0.1cm}

Massive gravity is obtained by
requiring that the space-time  geometry at $u=1$ is
flat. One  way to impose this boundary condition is
to require that $\t g_{\mu\nu}(x,u=1)=\eta_{\mu\nu}\equiv
{\rm diag}(-1,1,1,1)$  \cite{GG,CdR}.
Then, the auxiliary dimension can be  ``integrated out''.
This produces quadratic and nonlinear terms in the resulting
effective Lagrangian with specially tuned coefficients.  
In the quadratic order one gets
the FP term  \cite{GG,CdR}.  We will show below that the cubic
order result  reads as follows:
\beq
{\cal L} =  {\mpl^2 } \sqrt{g} R -   {\mpl^2 m^2 \over 4}
\left (  h^2_{\mu\nu} - h^2   -  h_\mu^\nu h_\nu^\alpha h_\alpha^\mu
+{5\over 4} hh_{\alpha \beta}^2  - {1\over 4} h^3 \right )\,,
\label{PFGen}
\eeq
where $h_{\mu\nu} \equiv g_{\mu\nu}-\eta_{\mu\nu}$; all the indices
in the mass term in (\ref {PFGen}) are  raised by $\eta^{\mu\nu}$, and
the  $\sqrt{g} R $ term in (\ref {PFGen})
should also be expanded up to the cubic order.

For small perturbations the quadratic FP term in
(\ref {PFGen}) describes five degrees of freedom of massive spin-2.
Naively, this would seem enough for perturbative
consistency. Regretfully, the issue is more subtle: to see a 
potential problem  one can consider a  weak field of a
locally-nontrivial asymptotically-flat  solution  -- say a
weak field of a static lump of matter (below we will refer
such backgrounds as {\it local}).  Then, the expression
for $h_{\mu\nu}$ can be decomposed as
\beq
h_{\mu\nu} = h^{cl}_{\mu\nu} + \delta_{\mu\nu},
\label{hcl}
\eeq
where  $h^{cl}_{\mu\nu}$ denotes the weak field of the local background
and $\delta_{\mu\nu}$ denotes small fluctuations  about it,
$1 \gg h^{cl}_{\mu\nu} \gg  \delta_{\mu\nu}$.
Substituting (\ref {hcl}) into (\ref {PFGen})
one gets  new quadratic terms for the fluctuations,
$ h^{cl} \delta \delta$.  Since  $h^{cl} \ll 1$ these new terms are
smaller than  the  $\delta \delta$ terms. Nevertheless,
they  may destroy the delicate balance of the FP term
and introduce a ghost.  In a generic nonlinear
completion  of  the FP massive gravity such a ghost does appear
as a sixth  degree of freedom \cite{BD}.

In order for the sixth mode not to emerge
a very special tuning  of the coefficients  of the quadratic and
cubic terms is needed, as was observed in \cite{AGS} and worked
out in \cite{Creminelli}. What is interesting,  is that the
theory  (\ref{egr}) automatically produces such
tuned coefficients up to the cubic order!

There could have been four arbitrary coefficients in front
of the five quadratic and cubic  order terms in   (\ref {PFGen}).
In the initial Lagrangian (\ref {egr}) we tuned the
coefficient between the two terms under the $u$-integral.
As a result,  we have  automatically generated four ``good''
coefficients in  (\ref {PFGen})\footnote{In general, there exists
a one-parameter family of cubic order terms for which the
sixth derivative terms for helicity-0 do cancel out,
$(h^2_{\mu\nu} - h^2   +c_1 h_\mu^\nu h_\nu^\alpha h_\alpha^\mu -
{6c_1+1 \over 4 }  hh_{\alpha \beta}^2  + {2c_1 +1\over  4} h^3) $.
The theory (\ref {egr}) with the boundary conditions used here
gives  $c_1=-1$. From this point of view,  (\ref {egr}) had to
generate only three ``good''  coefficients up to this order.}.
This hints towards  a hidden symmetry of (\ref {egr}) which may be
responsible  for such arrangements.

One goal of the present paper is to derive  the
terms presented in (\ref {PFGen}) from the
Lagrangian (\ref {egr}). Furthermore,
after observing the cancellation of the ``ghost terms'',
we will study the remaining nonlinear interactions
of the helicity $\pm2$, $\pm1$ and helicity-0 states.

Thus, we will be looking at the cubic order Lagrangian (\ref {PFGen})
in the decoupling limit where the nonlinear dynamics
of all the helicities  can be made manifest.
This method -- first used for massive non-Abelian gauge fields
in Ref. \cite{Khriplovich}, and  developed for massive gravity
in  \cite{AGS} --  proved to be successful in identifying
the presence or  absence  of the sixth degree of freedom in
nonlinear theories \cite{AGS,DeffayetRombouts,Creminelli,Rubakov}.
The limit we consider reads as follows:
\beq
m \to 0, ~~~\mpl \to \infty, ~~~  \Lambda_3\equiv  (m^2 \mpl )^{1/3}
 ~~{\rm is~fixed}\,.
\label{declim3}
\eeq
By taking this limit in  (\ref {PFGen}) we show that:
(I) The quadratic theory propagates
the helicity $\pm2$  modes described by the tensor
field ${\bar h}_{\mu\nu}$, the helicity $\pm1$ modes described by
the vector field  $A_\mu$, and  the helicity-0 mode described by
the scalar field $\pi$ -- all with canonical kinetic terms.
(II) The  cubic terms with six derivatives, $(\partial \partial \pi)^3$,
which would give rise to the sixth degree of freedom on local backgrounds,
cancel out automatically due to the  special values of the coefficients
of the quadratic and cubic terms in (\ref {PFGen}).
(III) There is a four-derivative cubic term for the helicity-0 mode,
$ \square \pi (\partial \pi)^2$,  which  was
first found in the context of the DGP model \cite{DGP} in Ref.
\cite{Ratt}. This term is  ghost-free for
local  backgrounds; it is also invariant under the
``galilean'' transformation in the $\pi$ space,
$\partial_\mu \pi  \to  \partial_\mu \pi+v_\mu $,
with $v_\mu $ being a constant four-vector.
(IV) There are
four-derivative cubic terms that mix two helicity-0 fields
with one helicity-2, such as ${\bar h}
((\square\pi)^2 - (\partial\partial \pi)^2)$,
or two helicity-1 fields  with one helicity-0, such as
$\partial \partial \pi (\partial A)^2$.
All of them are ``galilean''  invariant, and none of these terms gives
rise to ghosts on the local backgrounds.
(V) If the external
sources are ignored (or outside of localized sources) all the
cubic mixings between the helicity-2 and helicity-0, and
helicity-1 and helicity-0,   can be removed
by a nonlinear redefinition of  the helicity-2  and helicity-1 fields,
respectively. What  remains is the term,  $\square \pi (\partial \pi)^2$.
Interestingly, the above redefinition  of ${\bar h}_{\mu\nu}$
generates also the quartic Galileon  term,
$(\partial \pi)^2 ( (\square \pi)^2 - (\partial \partial \pi)^2 $
\cite{Galileon}.  The latter is known to be ghost-free \cite{Galileon}
(for more general studies of the Galileon, see, \cite{GalileonCedric}).

That the obtained terms contain the four-derivative cubic
term -- the cubic Galileon $\square \pi (\partial \pi)^2$
--  identical to the one  found in Ref. \cite{Ratt} in the context
of the DGP model, is a  hint:  The Galileon  terms emerge  in  theories with
spontaneously broken symmetries \cite{CdRAT}.
Hence, our findings suggest  that there should be
an underlying nonlinearly realized
symmetry of (\ref {egr}), which is responsible for
the cancellations of the ``ghost  terms''.
It is conceivable that this symmetry  is related to
a spontaneously broken  5D reparametrization invariance
hiddenly present in the model (\ref {egr}). This
symmetry should also be helpful in addressing the issue
of stability of (\ref {egr}) with respect to
quantum gravity loops, which has been
left open for now.

So far we have focused on the cubic order. How about an arbitrary
$n^\prime th$ order terms,  $(h^{cl})^{n-2}\delta \delta$?  Making
sure that the sixth degree
of freedom does not show up order-by-order would be a tedious
program\footnote{Since in each consecutive order these terms
have  smaller and smaller coefficients, one could try to show that
after a certain order the ghost appears only above a certain
UV cutoff of the low energy theory.}.
However, we are in a better position here as the
Lagrangian (\ref {egr}) sums up all the polynomial
terms in $h_{\mu\nu}$.  Then, the presence of a  ghost
could be seen by calculating the exact Hamiltonian. Boulware and Deser (BD)
\cite{BD}  have shown  that the sixth degree of freedom
would in general appear in nonlinear massive gravity
due to the loss of the Hamiltonian constraint:  This
leads to a Hamiltonian term that is proportional to
positive powers of the canonical momenta, but is sign indefinite,
hence, signaling the presence of a ghost.

Therefore, the absence of the BD term  would be  a good indicator that
the sixth degree of freedom is not present\footnote{It is not a guarantee, 
however, of the absence of other types of ghosts that may be present in 
a theory for some other reasons (e.g., introduced by hand). 
Here, we're focusing on the ghost that may appear 
in massive gravity  due to the nonlinear interactions.}. 
The Hamiltonian for
(\ref {egr}) was calculated in \cite{GG}, where it was shown that the BD
term cancels out.  Hence, one should expect that the BD ghost
does not appear in the order-by-order expansion of (\ref {egr}).
Moreover, in Ref. \cite{CdR}  the decoupling limit
of the theory was considered and 
it was shown to all orders that the leading terms arising 
at the scale $\Lambda<\Lambda_3$,  that could give rise to ghosts, 
cancel out\footnote{Ref. \cite{Creminelli} concluded that  no linear
combination of the quartic order terms in $h_{\mu\nu}$ can give a theory for
$\pi$ that would be ghost-free.  This issue will be revisited in our
forthcoming paper \cite{dRG},  with a different conclusion.
Until then,  we ignore the explicit quartic and higher
terms in $h_{\mu\nu}$.}.

There have been proposals in the literature to obtain  the theory
of massive spin-2 via a dynamical  condensation mechanism
(for recent works see, e.g.,  \cite{Zura,Slava} and references therein).
It would be interesting to see whether the cubic  terms in these models can
also automatically give rise to ghost-free structures for
the $\pi$ field.

\subsection*{2. Integrating out the auxiliary dimension}

The goal here is to calculate order-by-order the $u$-dependence of the
extended metric ${\tilde g}_{\mu\nu}(x, u)$, then substitute it back
into (\ref {egr}),  and integrate  the latter  w.r.t. $u$. This should give
the effective Lagrangian written  in terms of $h_{\mu\nu}$ only.
To fulfill  this goal we introduce the notations
\beq
{\tilde g}_{\mu\nu}(x, u) = \eta_{\mu\nu}+ H^{(1)}_{\mu\nu}(x,u)+
H^{(2)}_{\mu\nu}(x,u)+ H^{(3)}_{\mu\nu}(x,u)+...,
\label{gH}
\eeq
where $H^{(1)}_{\mu\nu}(x,u),~H^{(2)}_{\mu\nu}(x,u), ...$ are
perturbations in the corresponding order.
Since  ${\tilde g}_{\mu\nu}(x, u=0)=g_{\mu\nu}(x) =\eta_{\mu\nu}+
h_{\mu\nu}(x)$,  we require that
\beq
 H^{(1)}_{\mu\nu}(x,u=0)= h_{\mu\nu}(x),~~~ H^{(2)}_{\mu\nu}(x,u=0)=
H^{(3)}_{\mu\nu}(x,u=0)=0.
\label{Hbc}
\eeq
Furthermore, following Refs. \cite{GG,CdR} we impose the boundary condition
on the extended metric ${\tilde g}_{\mu\nu}(x, u=1)=\eta_{\mu\nu}$,
which guarantees that the theory reduces to  FP massive gravity
in the quadratic approximation. This boundary condition implies that
$H^{(1)}_{\mu\nu}(x,u=1)= H^{(2)}_{\mu\nu}(x,u=1)=
H^{(3)}_{\mu\nu}(x,u=1)=0$. Having set these, the calculation of
the $u$-dependence of ${\tilde g}_{\mu\nu}(x, u)$
is well defined.

The expression for the mass  term in (\ref {PFGen}) in terms of
$H^{(1)}_{\mu\nu}(x,u),~H^{(2)}_{\mu\nu}(x,u), ...$, up to
cubic  order,  is straightforward to obtain:
\ba
-{\mpl^2 m^2\over 2} 2 \int_{0}^{+1} {du \over 4}\Big (
 (\partial_u H^{(1)}_{\mu\nu})^2 + 2 \partial_u H^{(1)}_{\mu\nu}
\partial_u H^{(2)}_{\mu\nu} - 2  H^{(1)}_{\mu\nu} \partial_u
H^{(1)}_{\nu\alpha} \partial_u H^{(1)}_{\alpha \mu} -
(\partial_u H^{(1)})^2
\nonumber \\ - 2 \partial_u H^{(1)} \partial_u H^{(2)}
+ 2 \partial_u H^{(1)} H^{(1)}_{\mu\nu} \partial_u H^{(1)}_{\mu\nu}
+{1\over 2} H^{(1)} (\partial_u H^{(1)}_{\mu\nu})^2 -
{1\over 2} H^{(1)} (\partial_u H^{(1)})^2 \Big),
\label{umass}
\ea
where  we used the fact that  the metric is a
${\bf Z_2}$ symmetric function of $u$, and set the integration
limits from $0$ to $1$. Also, in (\ref {umass}) and in what follows,
we use simplified notations with all lower-case indices
contracted by the flat space-time metric.

To find the $u$-dependence of
$H^{(1)}_{\mu\nu}(x,u),~H^{(2)}_{\mu\nu}(x,u), ...$
we integrate the equations of motion. For this we vary
the action (\ref {egr}) w.r.t. ${\tilde g}_{\mu\nu}(x, u)$.
The resulting equations for $u=0^+$ and $0<u\leq 1$ read respectively
as follows:
\beq
G_{\mu\nu}  - m^2 \left (k_{\mu\nu} - g_{\mu\nu} k \right ) =
T_{\mu\nu}/(2\mpl^2)\,,
\label{junction}
\eeq
and
\beq
\partial_u
\left [ \sqrt{\tilde g} \left (k {\tilde g}^{\mu\nu} -
k^{\mu\nu} \right ) \right ] =  {1\over 2} {\tilde g}^{\mu\nu }\sqrt{\tilde g}
\left ( k^2  -   k_{\alpha\beta}^2   \right ) + 2
\sqrt{\tilde  g} \left ( k^{\mu\rho} k_{\rho}^{\nu} -k^{\mu\nu}k \right )\,.
\label{bulk}
\eeq
The latter is the equation  that determines the
``$u$-evolution''  of the extended metric.  With the boundary conditions
specified above it is straightforward to solve  (\ref {bulk})
for  $H^{(1)}_{\mu\nu}(x,u),~H^{(2)}_{\mu\nu}(x,u), ...$.
It turns out that only the solution for $H^{(1)}_{\mu\nu}(x,u)$ is needed at
the cubic order. The latter reads:
\beq
H^{(1)}_{\mu\nu}(x,u)= (1-u)h_{\mu\nu}(x).
\label{H1H2}
\eeq
This expression for  $H^{(1)}_{\mu\nu}(x,u)$ guarantees that
the two terms in (\ref {umass}),  which contain the function
$H^{(2)}_{\mu\nu}(x,u)$,  integrate to zero due to the
boundary conditions.  Substituting the expression (\ref {H1H2})
into (\ref {egr}) and integrating it w.r.t. $u$ we obtain
\beq
-   {\mpl^2 m^2 \over 4}
\left (  h^2_{\mu\nu} - h^2   -  h_\mu^\nu h_\nu^\alpha h_\alpha^\mu
+{5\over 4} hh_{\alpha \beta}^2  - {1\over 4} h^3 \right )\,,
\label{PFmass}
\eeq
which is the mass term presented earlier in (\ref {PFGen}).

\subsection*{3. Extracting the longitudinal mode}

We begin by rewriting the Lagrangian density
(\ref {PFGen}) in a manifestly covariant form. For this, following
Ref. \cite{AGS},  we introduce a covariant tensor field $H_{\mu\nu}$, which
is related to  $g_{\mu\nu} $ as follows:
\beq
H_{\mu\nu} = g_{\mu\nu} - \eta_{ab}{\partial \phi^a(x)\over \partial x^\mu}
{\partial \phi^b(x)\over \partial x^\nu}\,,
\label{Hg}
\eeq
where $\phi^a(x)$ are just four scalars;  $a,b=0,1,2,3$, and
$\eta_{ab}={\rm diag} (-1,1,1,1)$ is the flat metric on the
field space of the scalars. This construction guarantees that
$H_{\mu\nu}$ transforms as a covariant symmetric rank-2
tensor.

Furthermore, it is convenient to decompose the scalars
as  $\phi^a(x) = x^a-\pi^a(x)$, where
$x^a \equiv \delta^a_\mu x^\mu$. This decomposition specifies that
under the general coordinate transformations,
$x^\mu \to x^\mu+ \zeta^\mu(x)$, the fields  $\pi^a(x)$
transform as $\pi^a  \to \pi^a + \delta^a_\mu \zeta^\mu$.
Using the above definitions,  we can easily find the expression for
the tensor $H_{\mu\nu}$ in terms of
$h_{\mu\nu} \equiv {\tilde h}_{\mu\nu}/ \mpl $:
\beq
H_{\mu\nu} =  { {\tilde h}_{\mu\nu} \over \mpl} +
{\partial_\mu  V_\nu  + \partial_\nu  V_\mu \over \Lambda_3^3} -
{\partial_\mu  V_\alpha \partial_\nu  V_\alpha  \over
\Lambda_3^6} \,,
\label{H}
\eeq
where $V_\mu\equiv  \delta_\mu^ a \pi_a \Lambda_3^3$,  is a field
that shifts as $V_\mu \to V_\mu +\eta_{\mu\nu} \zeta^\nu \Lambda_3^3$
under the general coordinate transformations (the index 
contraction  in (\ref {H}) is  done with $\eta^{\mu\nu}$).

Then, the Lagrangian  density (\ref {PFGen}) can be written in terms of
the covariant tensors only. It reads as follows:
\beq
{\cal L} =  {\mpl^2 } \sqrt{g} R -   {\mpl^2 m^2 \over 4}
\sqrt{g} \left (  H^2_{\mu\nu} - H^2   + H_\mu^\nu H_\nu^\alpha H_\alpha^\mu
- {5\over 4} HH_{\alpha \beta}^2  + {1\over 4} H^3  \right )\,,
\label{PFGen1}
\eeq
where all the indices are raised with
the metric  tensor $g^{\mu\nu}$ (as a consequence, 
the signs in front of the cubic terms in (\ref {PFGen1}) flip  
as compared with (\ref {PFmass})).  
In the cubic order (\ref {PFGen1}) reduces to  (\ref {PFGen})
after using the gauge fixing condition $V_\mu=0$.

Moreover, the external sources/fields couple to $g_{\mu\nu}$:
For instance the Lagrangian for a scalar field $\psi$ coupled to gravity
would  read as follows:
$$\mathcal L_\Psi={1\over 2} \sqrt{g}(-g^{\mu\nu}\partial_\mu \psi
\partial_\nu \psi -  2 V(\psi)).$$
We will not write explicitly the couplings of  $g_{\mu\nu}$ to the
external sources/field below, but will keep them in mind
(see, discussions below).

To turn to the decoupling limit  we decompose the vector field
$V_\mu$ as follows:
\beq
V_\mu = mA_\mu + \partial_\mu \pi\,,
\label{VA}
\eeq
where both $A_\mu$  and $\pi$ are kept
finite in the limit (\ref {declim3}).
The field $A_\mu$ will end up encoding the
helicity $\pm 1$ states,  while the field $\pi$ will
describe the helicity-0 state.

\vspace{0.1cm}

Next we calculate  the decoupling limit of the Lagrangian
(\ref {PFGen1}). This is done by
substituting (\ref {H}) and (\ref {VA})  into   (\ref {PFGen1}),
and taking the limit (\ref {declim3}). This procedure -- valid for
fields that decay fast enough at spatial infinity --
requires some care: we introduce an   infrared regulator of the theory,
say a large sphere of radius $L \gg 1/m$, and  take the radius
to infinity,  $L\to \infty $, before taking the limit (\ref {declim3}).
This hierarchy of scales enables us to put all the surface
terms to zero before taking the decoupling limit.

Once the above  procedure is adopted we find the following remarkable
properties: (1) All the terms containing six derivatives and
three helicity-0 fields, such as $ (\partial^2 \pi)^3$, that
come suppressed by the scale $\Lambda_5 \equiv (\mpl m^4)^{1/5}\ll
\Lambda_3$, cancel out up to a total derivative \cite{CdR}.
(2) The quadratic terms in $A$  form the Maxwell term,
while all the terms that are
linear in $A$ and quadratic in $\pi$,
such as $\partial A \partial^2 \pi \partial^2 \pi $,  which
would be suppressed by the scale $\Lambda_4 \equiv
(\mpl m^3)^{1/4}\ll\Lambda_3$, also cancel out
up to a total derivative.  (3) The only terms
that survive are those suppressed by the scale  $\Lambda_3$.

In this section we focus only on the helicity-2 and helicity-0
modes,  while the terms with the helicity-1 field  will be ignored
until  the next section,  where they are shown to be harmless.

The remaining terms, after the conformal
transformation ${\tilde h}_{\mu\nu}={\bar h}_{\mu\nu}+
\eta_{\mu\nu}\pi$ that diagonalizes the quadratic action,  
read as follows:
\ba
{\cal L}^{\rm lim}_{\Lambda_3}&=&
-{1\over 2} {\bar h}_{\mu\nu} {\cal E}^{\mu\nu\alpha\beta}
{\bar h}_{\alpha \beta} + {3 \over 2} \pi \square \pi   +
{3 \square \pi  (\partial_\mu  \pi)^2 \over 4 \Lambda_3^3}
\nonumber \\
&&+ { {\bar h}\left ((\partial_\mu\partial_\nu \pi)^2 -  (\square \pi)^2
\right ) \over 4 \Lambda_3^3}  +   { {\bar  h}_{\mu\nu}
\left (  \partial_\mu\partial_\nu \pi \square \pi  -
\partial_\mu \partial_\alpha \pi  \partial_\nu \partial_\alpha \pi
\right ) \over 2 \Lambda_3^3} .
\label{LimH}
\ea
Here, all the indices are raised using the flat space metric and we do not
distinguish between the upper and lower cases.
${\cal E}$ denotes the Einstein operator that is related to the
linearized Einstein tensor $G_{\mu\nu}$ as follows:
$${\cal E}^{\mu\nu\alpha\beta} {\bar h}_{\alpha \beta}= G^{\mu\nu}
=- {1\over 2} ( \square {\bar h}^{\mu \nu} - \partial^\mu \partial^\alpha
{\bar h}_{\alpha}^{\nu} - \partial^\nu \partial^\alpha
{\bar h}_{\alpha}^{\mu} + \partial^\mu \partial^\nu {\bar h} -
\eta^{\mu\nu} \square {\bar h} + \eta^{\mu\nu}
\partial_\alpha  \partial_\beta {\bar h}^{\alpha\beta}).$$
All the terms in  (\ref {LimH}), up to total derivatives,
are invariant under  the ``galilean'' transformations of the $\pi$ field,
$\partial_\mu \pi \to \partial_\mu \pi +v_\mu$,
where $v_\mu$ is some constant four-vector.
Moreover,  none of the  cubic terms  in (\ref {LimH})
produce ghosts on  any local background.
Indeed, the  last term in the first line of
(\ref {LimH}) is identical to the one  found in the decoupling limit
of the DGP model \cite{Ratt} (see, also \cite{GI} for related discussions).
This term is known to give rise to the equations of
motion that have no more than  two derivatives acting on each
fields \cite{Ratt}.
To see that the rest of the terms in (\ref {LimH})  are also safe
we rewrite them as follows:
\beq
{\bar h} \left (  (\partial_\mu\partial_\nu \pi)^2 -  (\square \pi)^2
\right )= {\bar h} \left ( 2 \partial_0^2 \pi \Delta \pi -2 (\partial_0
\partial_j \pi)^2 - (\Delta \pi)^2 + (\partial_i
\partial_j \pi)^2    \right )\,,
\label{O31}
\eeq
and
\beq
{\bar  h}_{\mu\nu}
\left (  \partial_\mu\partial_\nu \pi \square \pi  -
\partial_\mu \partial_\alpha \pi  \partial_\nu \partial_\alpha \pi
\right ) = {\bar  h}_{00}  \left (  \partial_0^2 \pi  \Delta \pi -  (\partial_0
\partial_j \pi)^2 \right )+ \nonumber \\
 2 {\bar  h}_{0j} \left (
 \partial_0 \partial_j \pi  \Delta \pi  - \partial_0  \partial_k  \pi
\partial_j  \partial_k  \pi \right )+ {\bar  h}_{ij}
\left ( \partial_i  \partial_j \pi \square \pi - \partial_i  \partial_\nu
\pi  \partial_j  \partial_\nu \pi \right )\,.
\label{O32}
\eeq
These  have at most two time derivatives, and will not  produce
any terms with more than two time derivatives in the equations of motion.
Importantly, the term multiplying ${\bar  h}_{00}$ in 
(\ref {LimH}) has no time derivatives, only spatial ones, showing that 
${\bar  h}_{00}$ remains a Lagrange multiplier at the cubic order in the 
decoupling limit.

The expression (\ref {LimH}) is invariant under
the gauge transformations, $\delta {\bar h}_{\mu\nu} =-
\partial_\mu \zeta_\nu - \partial_\nu \zeta_\mu$,
due to  the cancellation  between
the two nonlinear terms in the second line.
In other words,  the Bianchi identity  is
automatically  satisfied  for the tensor equation that follows
from (\ref {LimH}) by varying it w.r.t.
${\bar h}_{\mu\nu}$.

If  we ignore coupling to external sources, one can simplify
further  the Lagrangian (\ref {LimH}) by the following
nonlinear transformation
\beq
{\bar h}_{\mu\nu} =  {\bar h}^\prime_{\mu\nu} +  {1\over 2 \Lambda_3^3}
\partial_\mu\pi \partial_\nu \pi\,.
\label{NLdiag}
\eeq
The resulting Lagrangian reads:
\beq
{\cal L}^{lim}_{\Lambda_3}=
-{1\over 2} {\bar h}^\prime_{\mu\nu} {\cal E}^{\mu\nu\alpha\beta}
{\bar h}^\prime_{\alpha \beta} + {3 \over 2} \pi \square \pi   +
{3 \square \pi  (\partial_\mu  \pi)^2 \over 4 \Lambda_3^3}+...
\label{DGPphi}
\eeq
As emphasized before, in the present theory  it is the field $g_{\mu\nu}$
that couples to external sources/fields.
In the linearized theory the linearized source/field
stress-tensor $T_{\mu\nu}$ couples  as  ${\tilde h}_{\mu\nu}T_{\mu\nu}/\mpl$,
which after the  conformal transformation  reads as follows
$({\bar h}^\prime_{\mu\nu}T_{\mu\nu} + \pi T)/\mpl$. These couplings, e.g.,
for a static source,  are held fixed and finite in the decoupling
limit (i.e. $ T_{\mu\nu}/\mpl$  is fixed to be finite \cite{Ratt}).
However, the diagonalization  of the  nonlinear terms performed by
(\ref {NLdiag}) would generate an additional  nonlinear coupling
$\partial_\mu\pi \partial_\nu \pi T_{\mu\nu}/(\mpl \Lambda_3^3)$.
Hence, to avoid complications with the additional nonlinear couplings
and the modified light-cone, it is better to think of the ``decoupling'' 
limit Lagrangian (\ref {LimH}) in which the helicity-2 and helicity-0 field
mix at the cubic order.  For certain sources -- such as
static ones --  the additional coupling
$\partial_\mu\pi \partial_\nu \pi T_{\mu\nu}/(\mpl \Lambda_3^3)$
is  zero, and therefore  using the Lagrangian
(\ref {DGPphi}) in which  the helicity-2 and helicity-0
are truly decoupled may be more convenient.

Interestingly, the field redefinition (\ref {NLdiag}) in
(\ref {LimH})   generates in (\ref {DGPphi}) the term
\beq
{2 \partial_\mu \pi \partial_\nu \pi ( \partial_\mu \partial_\beta \pi
 \partial_\nu \partial_\beta \pi - \partial_\mu \partial_\nu \pi
\square \pi) +
(\partial_\mu \pi)^2 ( (\square \pi)^2 -
(\partial_\alpha \partial_\beta \pi)^2)\over
16 \Lambda_6^3}\,.
\label{quart}
\eeq
This is exactly the  quartic Galileon
introduced in \cite{Galileon} as a ghost-free quartic order term
giving rise to two derivative equations of motion.
Note that the coefficients of the cubic Galileon in (\ref {DGPphi}) and
the quartic Galileon  in (\ref {quart}) are related to each other
since they both originate in the cubic mixing terms
of helicity-2 with  helicity-0 in (\ref {LimH}). In the present paper
we ignore the quartic order terms in $h_{\mu\nu}$,
but in general,  depending on the coefficients of these terms,
the quartic Galileon  (\ref {quart}) may or may not cancel.
These issues will be discussed in detail in \cite{dRG}, where it will be shown
that in a general quartic-order theory, as soon as the cubic Galileon is
present in (\ref {DGPphi}), we are also bound to generate either the 
quartic Galileon,  or a quartic mixing,  or even both together accompanied
by the quintic Galileon.

In practice, the quartic  and higher order
terms  are  negligible at large scales where the tensor-scalar
gravity sets in \cite{vDVZ}, while they become as important as
the quadratic and cubic contributions around the Vainshtein scale
\cite{Arkady}  (see also \cite{DDGV}). The mixing terms will be essential 
to address  the issue whether  the full theory admits 
superluminal propagation,  as the pure Galileon terms do,
(see Refs.~\cite{Superluminal,Galileon}).

\subsection*{4. Extracting the helicity-1 modes}

The Lagrangian for the vector field in the decoupling limit
is obtained by substituting (\ref {H}) and (\ref {VA})
into   (\ref {PFGen1}), and taking the limit (\ref {declim3}).

As mentioned before, the quadratic terms in $A$  form the Maxwell term.
Moreover,  terms  linear in $A$ and quadratic in $\pi$, ({\it e.g.},
$\partial A \partial^2 \pi \partial^2 \pi $),
that would be suppressed by the scale $\Lambda_4 \equiv
(\mpl m^3)^{1/4}\ll\Lambda_3$, cancel out
up to  total derivatives.  The terms
that survive in the decoupling limit are
suppressed by the scale  $\Lambda_3$.
These mix  two helicity-1 fields with
helicity-0, $\partial A \partial A \partial^2 \pi$.
The resulting Lagrangian reads as follows:
\ba
{\cal L}_A = -{1\over 4} F_{\mu\nu}^2 & -& {\partial_\mu \partial_\nu \pi
\over 4 \Lambda_3^3} \( 2  \partial_\mu A_\alpha \partial_\nu
A_\alpha +2  \partial_\alpha  A_\mu \partial_\alpha  A_\nu
+ 8 \partial_\mu A_\alpha \partial_\alpha A_\nu - 12
\partial_\alpha A_\alpha  \partial_\mu A_\nu \) \nonumber \\
&-& {\square  \pi
\over 4 \Lambda_3^3} \( -  (\partial_\alpha A_\beta)^2 - 5
( \partial_\mu A_\nu \partial_\nu A_\mu) +
6 (\partial_\alpha A_\alpha)^2 \)\,.
\label{Vector}
\ea
As before, all the contractions are by $\eta_{\mu\nu}$,
and no distinction is made between the lower
and upper-case indices.

The above expression is invariant under
the  internal galilean transformations. However, the gauge invariance of
(\ref {Vector}) is not immediately obvious.  We leave it to
reader's pleasure to show that (\ref {Vector})
reduces, up to a total derivative, to the following Lagrangian:
\beq
{\cal L}_A = -{1\over 4} F_{\mu\nu}^2 - {1 \over 2 \Lambda_3^3}
A_\mu (\partial_\mu \partial_\nu
-\eta_{\mu\nu} \square) (\partial_\alpha  \pi F_{\alpha\nu})\,.
\label{Vector1}
\eeq
The above expression
is  invariant, up to a total derivative,  w.r.t. the gauge transformations of
the vector field, $A_\mu \to A_\mu+ \partial_\mu \chi $, where $\chi $
is the gauge parameter.

Furthermore,  the nonlinear terms in (\ref {Vector1})
can be removed by the following non-linear field redefinition:
\beq
A_\mu \to A_\mu - {1\over 2\Lambda_3^3} (\partial_\alpha  \pi F_{\alpha\mu})\,.
\label{AB}
\eeq
After this transformation, and up to quartic terms, we are left with
the Maxwell Lagrangian.  Therefore, (\ref {Vector}) describes helicity
$\pm1$ modes, and does not give  rise to ghosts at cubic order
on any local background.

\subsection*{5. Brief comments}

Since the summary of our main results has already been
given in Section 1, we  end  this work with  a few 
technical comments.

(i) In Section 3  we calculate the decoupling limit of
(\ref {PFGen1})  using the  method of Ref.~\cite{AGS}.
The method of Ref.~\cite{Rubakov},  although  similar to that of
\cite{AGS}, differs from it slightly and follows more
closely  the St\"uckelberg method for the gauge fields.
Furthermore, we have checked  that in the approach of  Ref.~\cite{Rubakov}
all the six-derivative cubic terms cancel out, and four derivative
ones remain. The mixing terms for helicity-2 and helicity-0
are present,  and take a somewhat different form, but satisfy gauge
invariance and yield automatically the Bianchi identities.  
One can also show that
those mixing terms are reducible by a nonlinear field redefinition to
the ones we obtained here, and are  also removable by yet another
field redefinition (if external fields/sources are ignored) at the expense
of generating the higher order terms.

(ii) It is straightforward to see that the special
coefficients in (\ref {PFGen}) also play  a role in the construction
of the Hamiltonian. Indeed, introducing the standard ADM decomposition
\cite{ADM}  with the lapse $N$ and shift $N_j$ we find that
the cubic order Hamiltonian  of (\ref {PFGen}) is
linear in $\delta N = N-1$.  Hence the Hamiltonian constraint is
maintained in this order\footnote{Note that $h_{00}$ enters quadratically,
but this does not prevent the Hamiltonian to be linear in $\delta N$,
which is the right variable at the nonlinear level and 
away from the decoupling limit.}.
On the other hand,  $N_j$  enters quadratically and is
algebraically determined, as required for massive spin-2.
Note that the decoupling limit considerations do not yet
guarantee  positive semidefiniteness of the Hamiltonian 
of the full  theory, and this has to be addressed 
separately (see discussions in \cite {GG}).

(iii) Ref.~\cite{Creminelli} showed that $\delta N$ does
get quadratic terms in the quartic order of a general massive
theory (see also \cite {BD,GGruzinov}).  This dependence, however, 
may come in a special combination with $N^2_j$ that allows to 
preserve the  Hamiltonian constraint and avoid the BD term, 
see  more on this in Ref. \cite{dRG}.

\subsubsection*{Acknowledgments}

We would like to thank Massimo Porrati for collaboration at an
early stage of the work,  and useful discussions.
We thank Giga Chkareuli, Justin Khoury,
David Pirtskhalava, Oriol Pujolas and  Itay Yavin
for useful conversations. GG was supported by NSF
grant PHY-0758032.  CdR was supported by the
Swiss National Foundation.

\end{document}